\documentclass[9pt,twocolumn,twoside]{osajnl}

\journal{optica} 

\usepackage{xcolor}
\usepackage{multirow}

\setboolean{shortarticle}{true}

\title{Michelson Holography: Dual-SLM Holography with Camera-in-the-loop Optimization}
\author[1,2]{Suyeon Choi}
\author[1,2,*]{Jonghyun Kim}
\author[2]{Yifan Peng}
\author[2]{Gordon Wetzstein}

\affil[1]{NVIDIA, 2788 San Tomas Expressway, Santa Clara, CA 95051, USA}
\affil[2]{Electrical Engineering Department, Stanford University, Stanford, California 94305, USA}

\affil[*]{Corresponding author: jonghyunk@nvidia.com}

\begin{abstract}
We introduce Michelson Holography (MH), a holographic display technology that optimizes image quality for emerging holographic near-eye displays. Using two spatial light modulators, MH is capable of leveraging destructive interference to optically cancel out undiffracted light corrupting the observed image. We calibrate this system using emerging camera-in-the-loop holography techniques and demonstrate state-of-the-art holographic 2D image quality. 
\end{abstract}

\setboolean{displaycopyright}{true}
\newcommand{\citl}{\textsc{citl}}

\newcommand{\prop}{g}
\newcommand{\firstprop}{\prop_1}
\newcommand{\secprop}{\prop_2}
\newcommand{\prophat}{\widehat{g}}
\newcommand{\firstprophat}{\prophat_1}
\newcommand{\secprophat}{\prophat_2}
\newcommand{\target}{a_{target}}

\newcommand{\efficiency}{\eta}
\newcommand{\fourier}{\mathcal{F}}
\newcommand{\loss}{\mathcal{L}}
\newcommand{\transfer}{\mathcal{H}}
\newcommand{\fx}{f_x}
\newcommand{\fy}{f_y}
\newcommand{\numpixelx}{N_x}
\newcommand{\numpixely}{N_y}
\begin{document}

\maketitle

Near-eye displays in virtual and augmented reality systems should deliver high-quality imagery while supporting focus cues, a large field of view, and a reasonably sized eye box within a compact device form factor. Holographic near-eye displays promise to deliver these capabilities and have made remarkable progress over the last few years~\cite{Chen:15, Li:16, Maimone:2017, yu2017ultrahigh, Shi:2017, Chakravarthula:2019, Park:2019, Padmanaban:2019, lee2020wide, kuo2020, smalley2013anisotropic}. Unlike conventional displays, however, holographic displays using phase-only spatial light modulators (SLMs) can only indirectly show an image by shaping a wave field such that the target image is created through interference. This has been a challenging problem for decades and the image quality achieved by computer-generated holography (CGH) is often significantly lower than that of conventional technology, such as liquid crystal displays. 

A fundamental challenge of phase-only SLMs used for holography is their low diffraction efficiency. Caused by their limited pixel fill factors, backplane architectures, and other factors, as much 20\% of incident light may not be diffracted~\cite{ronzitti2012lcos}. This creates the zeroth diffraction order, which typically interferes with the user-controlled diffraction orders and significantly degrades observed image quality.


In optics, on-axis and off-axis filtering schemes are the two most common techniques to minimize the zeroth diffraction order. On-axis filtering physically blocks the undiffracted beam at the Fourier plane, which inevitably also blocks some amount of the diffracted light that contributes to the low-frequency content. In addition, this blocking process is much more challenging when multiplexing three colors~\cite{cho2018dc}. 
Off-axis methods suffer from reduced field of view (using half of the first diffraction order) or lowered efficiency (using higher diffraction orders)---both of these factors are crucial for near-eye displays~\cite{kim2014anamorphic}. In addition, approaches that compensate for the zeroth-order beam have been described that model either a correction beam~\cite{palima2007holographic} or the pixelated structure of the SLM~\cite{liang2012suppression, improso2017suppression}. Most recently, camera-in-the-loop (\textsc{citl}) holography techniques have been described that can partially compensate the undiffracted light of an SLM using its diffracted component without having to explicitly model all of these terms~\cite{Peng2020NeuralHolography,Chakravarthula:2020}. However, none of these software-only approaches accounts for the physical limitations of the actual SLM, limiting the degree to which destructive interference can be utilized to cancel out the zeroth order.

\begin{figure}[t!]
\centering
\includegraphics[width=\linewidth]{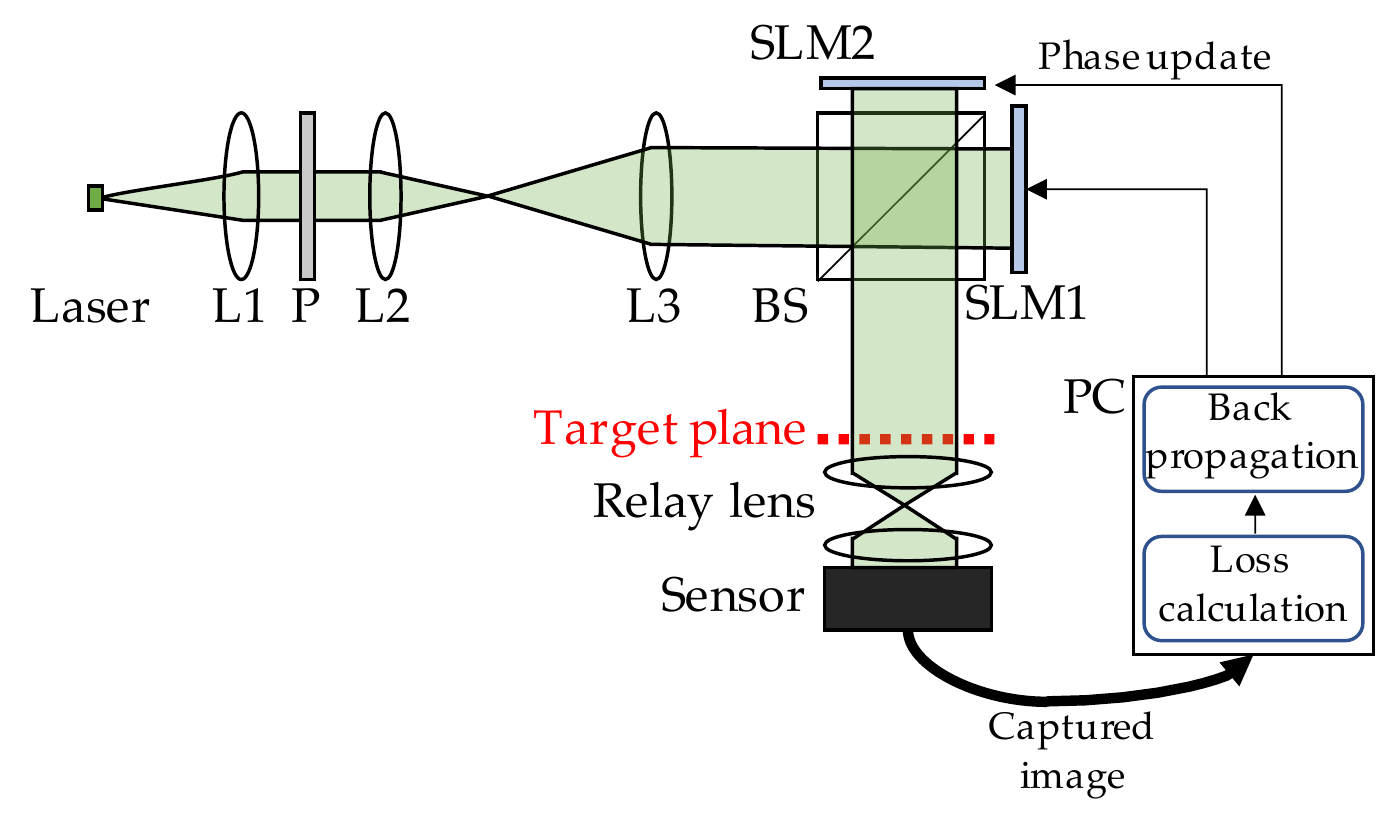}
\caption{Principle of Michelson Holography. L1: collimating lens, L2, L3: beam expanders, P: polarizer, BS: beam splitter. The sensor captures the intensity of the field at the target plane. The error between captured and target images is backpropagated, with a stochastic gradient descent optimization algorithm, to update the phase patterns of both SLMs.}
\label{fig:Principles}
\end{figure}




\begin{figure*}[t!]
\centering
\includegraphics[width=\linewidth]{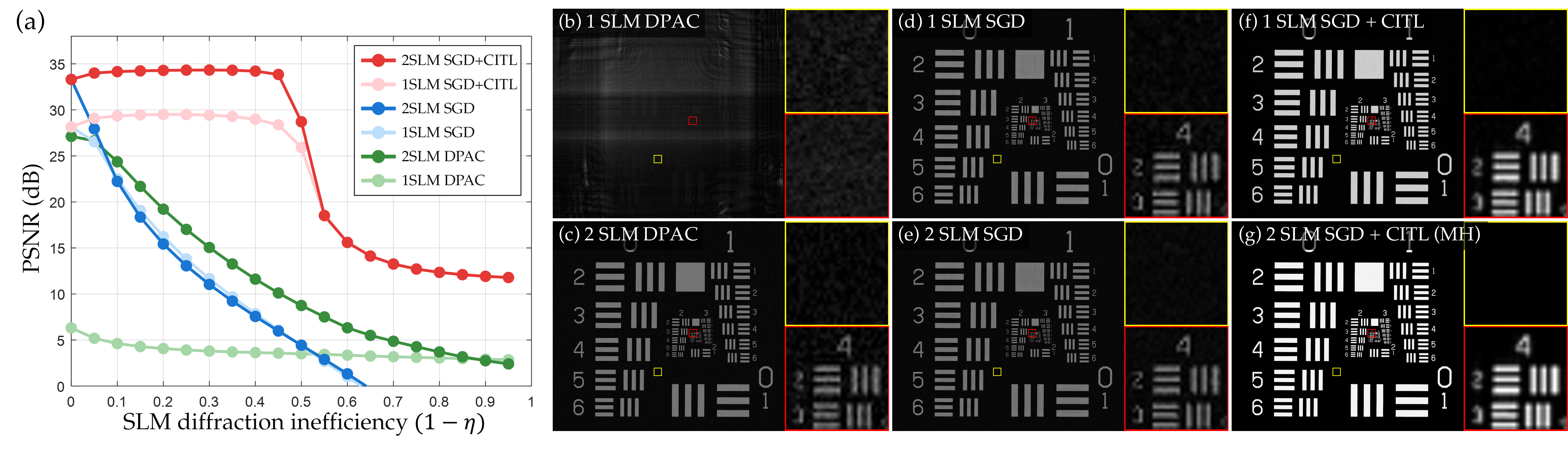}
\caption{Comparison of image quality achieved by several CGH techniques for varying amounts of SLM diffraction efficiency in simulation. (a) Quantitative evaluation of resolution chart. We report peak signal-to-noise ratio (PSNR). Double phase--amplitude coding (DPAC) with two SLMs works very well for high diffraction efficiencies, but the quality of DPAC quickly drops 
as the amount of undiffracted light increases. Similar trends are observed for the stochastic gradient descent (SGD) method. Using SGD with our \citl{} calibration method achieves high PSNR values for diffraction efficiencies as low as 50\%. (b-g) The reconstructed images, here shown for $1-\eta=0.2$, show similar trends, with Michelson Holography achieving the highest image quality and contrast. }
\label{fig:simulation_result}
\end{figure*}

Inspired by the design of Michelson interferometers, we propose Michelson Holography (MH), a holographic display architecture that uses two phase-only SLMs and a newly developed \textsc{citl} optimization procedure. The core idea of MH is to destructively interfere the diffracted light of one SLM with the undiffracted light of the other. As such, the undiffracted light can contribute to forming the target image rather than creating speckle and other artifacts. Our custom \textsc{citl} optimization procedure captures the coherent superposition of all diffracted and undiffracted light of the display and backpropagates the error w.r.t. a target image into both SLM patterns simultaneously. This procedure does not require us to explicitly model the SLM pixel structure or the undiffracted light. We simply need a camera that captures intermediate images of this iterative CGH algorithm, which automatically optimizes the resulting phase patterns for both SLMs. Although holographic displays with multiple SLMs have been investigated for viewing angle enhancement~\cite{Hahn:08}, improving resolution~\cite{yaracs2010multi}, or complex modulation~\cite{juday1991full, park2017characteristics, Shi:2017, macfaden2017characterization}, our approach is the first to leverage \textsc{citl} optimization to optimize image quality of two phase-only SLMs by mitigating the effect of undiffracted light in a fully automatic manner.

Figure~\ref{fig:Principles} shows the principle of MH. Here, a source field $u_{src}$ generated by the laser is incident on both SLMs. Each SLM delays the phase of the incident light by $\phi_{1/2}$, respectively, with diffraction efficiency $\efficiency$. The fields reflected off of the SLMs are $u_{1/2} = [\efficiency \exp^{i \phi_{1/2} (x,y)} + (1-\efficiency)] u_{src}$. The functions $\prop_{1/2}$ describe the unknown physical wave propagation from the SLMs to the target plane, including aberrations, whereas $\prophat_{1/2}$ describe their idealized models used in simulation. 
We use the angular spectrum method to model free-space wave propagation as
\begin{equation}
\begin{split}
&\prophat_{1/2}(\phi_{1/2}) = \iint \fourier (u_{1/2}) \ \transfer(\fx, \fy) \ e^{i2\pi(\fx x+\fy y)} \ d\fx d\fy, \\
&\transfer(\fx, \fy) = \begin{cases}
    e^{i\frac{2\pi}{\lambda}\sqrt{1- (\lambda \fx)^2- (\lambda \fy)^2}z}, & \text{if} \sqrt{\fx^2+\fy^2} < \frac{1}{\lambda},\\
    0 & \text{otherwise}.
  \end{cases}
\end{split}
\label{eq:asm}
\end{equation}
where $\mathcal{F}(\cdot)$ denotes Fourier transform, $\fx, \fy$ are spatial frequencies, $\lambda$ is the wavelength, $z$ is the propagation distance between SLM and target plane, and $x, y$ are the 2D coordinates on the SLM plane. In practice, the pixelated SLMs modulate the discrete phase patterns $\phi_{1/2} \in \mathbb{R}^{\numpixelx \times \numpixely}$, where $\numpixelx \times \numpixely$ is the SLM resolution, and $\prop_{1/2}, \prophat_{1/2} : \mathbb{C}^{\numpixelx \times \numpixely} \rightarrow \mathbb{C}^{\numpixelx\times \numpixely}$. 

Therefore, optimizing the two SLM phase patterns in MH is done by solving the problem
\begin{equation}
\underset{\phi_1, \phi_2}{\text{minimize}} \ \loss \left( s\cdot |\firstprophat(\phi_1) +  \secprophat(\phi_2)|, \target \right),
\label{eq:optimization_problem}
\end{equation}
where $\loss$ is a loss function, such as mean squared error, $\target$ is the amplitude of the target image, and $s$ is a scale factor. 

Most conventional CGH algorithms cannot easily model the proposed dual-SLM setup. Only the double phase--amplitude coding (DPAC) method~\cite{Hsueh:1978} is specifically designed for this case, because it represents a complex-valued field at the SLM plane $u=a e^{i\phi}$ as the sum of two phase-only fields with phases $\phi-\cos{^{-1}a}$ and $\phi+\cos{^{-1}a}$, respectively. Compared to the interlaced single-SLM DPAC variant~\cite{Maimone:2017}, dual-SLM DPAC is expected to achieve more accurate results, which is why we use it as a baseline for our algorithm. Note that DPAC does not account for undiffracted light, optical aberrations, or SLM misalignment. 

We build on the recently proposed \citl{} calibration approach~\cite{Peng2020NeuralHolography} and adopt it to our dual-SLM setup. This approach builds on a stochastic gradient descent (SGD) solution for Eq.~(\ref{eq:optimization_problem}). Starting from some initialization or a previous estimate of the two phase patterns $\phi_{1/2}^{(k-1)}$, we display these patterns on the SLM, capture an image on the target plane with a camera, use that to evaluate the loss function $\loss$, and then backpropagate the error using the gradients $\frac{\partial \loss}{\partial \phi_{1/2}}$ to find the next estimates $\phi_{1/2}^{(k)}$. Because we use a recorded image, this includes both diffracted and undiffracted contributions of both SLMs. Specifically, the update rules of our SGD-based solver in iteration $k$ are
\begin{equation}
\begin{split}
\phi_{1/2}^{(k)} \approx \phi_{1/2}^{(k-1)} & - \alpha \left( \frac{\partial \loss}{\partial \prop_{1/2}}\cdot\frac{\partial \prophat_{1/2}}{\partial \phi_{1/2}} \right)^T \times \\ &\loss \left( s\cdot |\firstprop\left(\phi_1^{(k-1)}\right) + \secprop\left(\phi_2^{(k-1)}\right)|, \target \right).
\label{eq:sgd_algorithm}
\end{split}
\end{equation}
%

We evaluate several CGH algorithms, including single- and dual-SLM variants of DPAC and SGD with and without \citl. Figure~\ref{fig:simulation_result} shows simulated results of a resolution chart at the target plane for varying SLM diffraction inefficiencies $(1-\efficiency)$. We simulate 1080p phase-only SLMs with 6.4~$\mu$m pixel pitch and $z=10$~cm propagation distance between SLM and target plane. In most cases, the peak signal-to-noise ratio (PSNR) rapidly decreases as the undiffracted light increases; only our SGD method with \citl{} optimization achieves a constant and high image quality up to $\approx 50$\% inefficiency for both single and dual-SLM setups. The proposed dual-SLM setup provides the best image quality of all methods, especially in the presence of undiffracted light. Figs.~ \ref{fig:simulation_result}(b)-(g) show qualitative comparisons of a reconstructed resolution target for $(1-\efficiency)=0.2$. These simulations demonstrate that MH achieves the best image contrast and quality. 


Our holographic display prototype uses two phase-only SLMs (HOLOEYE LETO and PLUTO) with 6.4~$\mu$m and 8.0~$\mu$m pixel pitch, respectively. The laser is a FISBA RGBeam fiber-coupled module with three optically aligned laser diodes with wavelengths 638~nm, 520~nm, and 450~nm. All results and the images for the \citl{} algorithm were captured with a FLIR Grasshopper3 2.3~MP color USB3 vision sensor through two Nikon AF-S Nikkor 35mm f/1.8 relay lenses. The \citl{} optimization was run on each color channel separately and full-color results were combined in post-processing. Additional hardware and software details are discussed in Supplement 1.

\begin{figure}[h!]
\centering
\includegraphics[width=\linewidth]{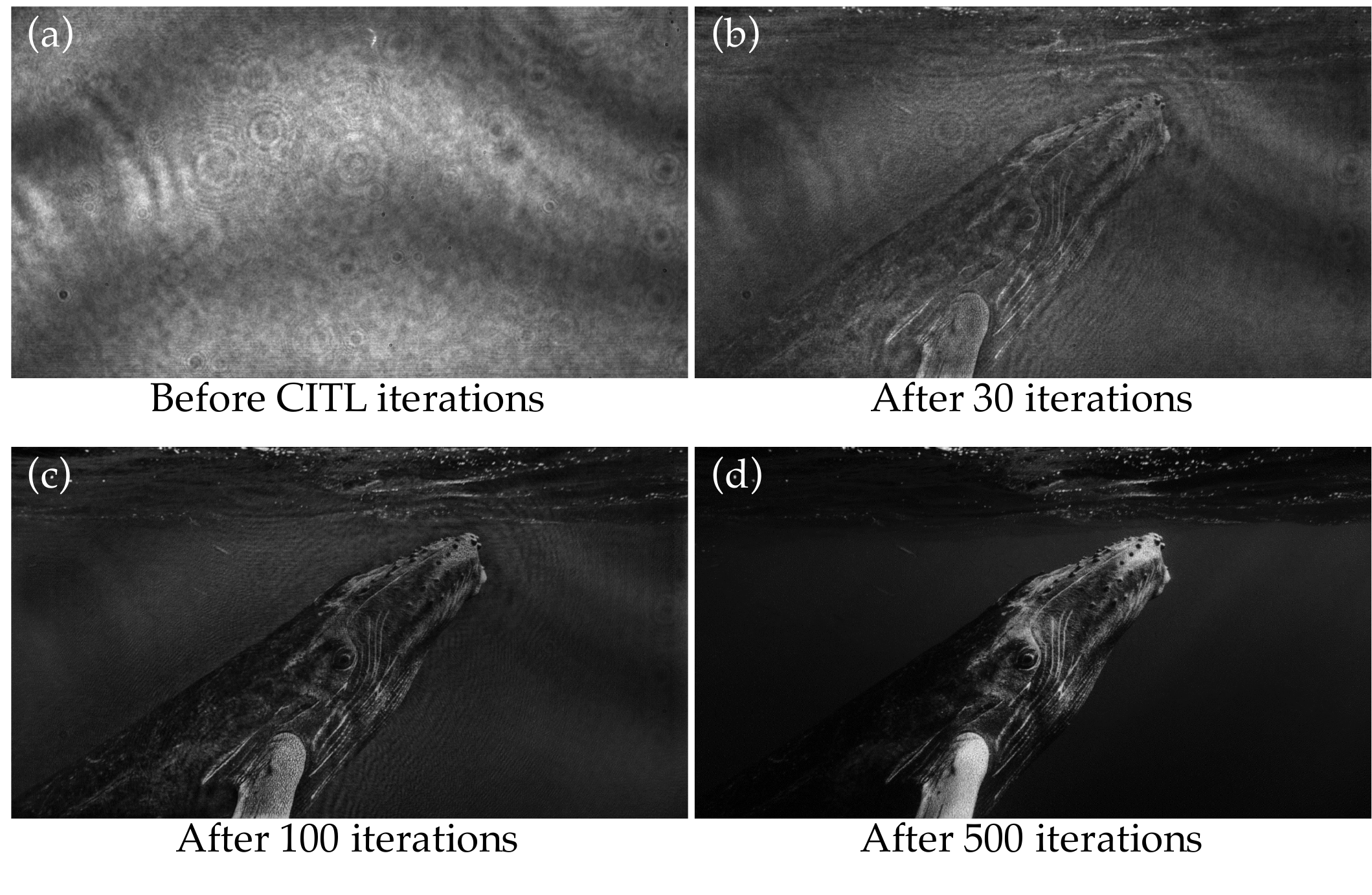}
\caption{Convergence of experimental results. (a) The undiffracted light from two SLMs naturally creates a fringe pattern. Our \citl{} optimization iteratively improves the observed image, as shown for (b) 30, (c) 100, and (d) 500 iterations (Visualization~1).}
\label{fig:fringe_pattern}
\end{figure}

Similar to a Michelson interferometer, the two SLMs of MH create fringe patterns on the target plane when no pattern is displayed (Fig.~\ref{fig:fringe_pattern}(a)). This is what makes calibrating a dual-SLM setup so challenging, but it is also this interference
that gives MH the ability to produce higher-contrast images than single-SLM setups. Using the proposed \citl{} optimization, the target image is found after about 500 iterations (Figs.~\ref{fig:fringe_pattern}(b)-(d), Visualization~1).


\begin{figure*}[t!]
\centering
\includegraphics[width=\linewidth]{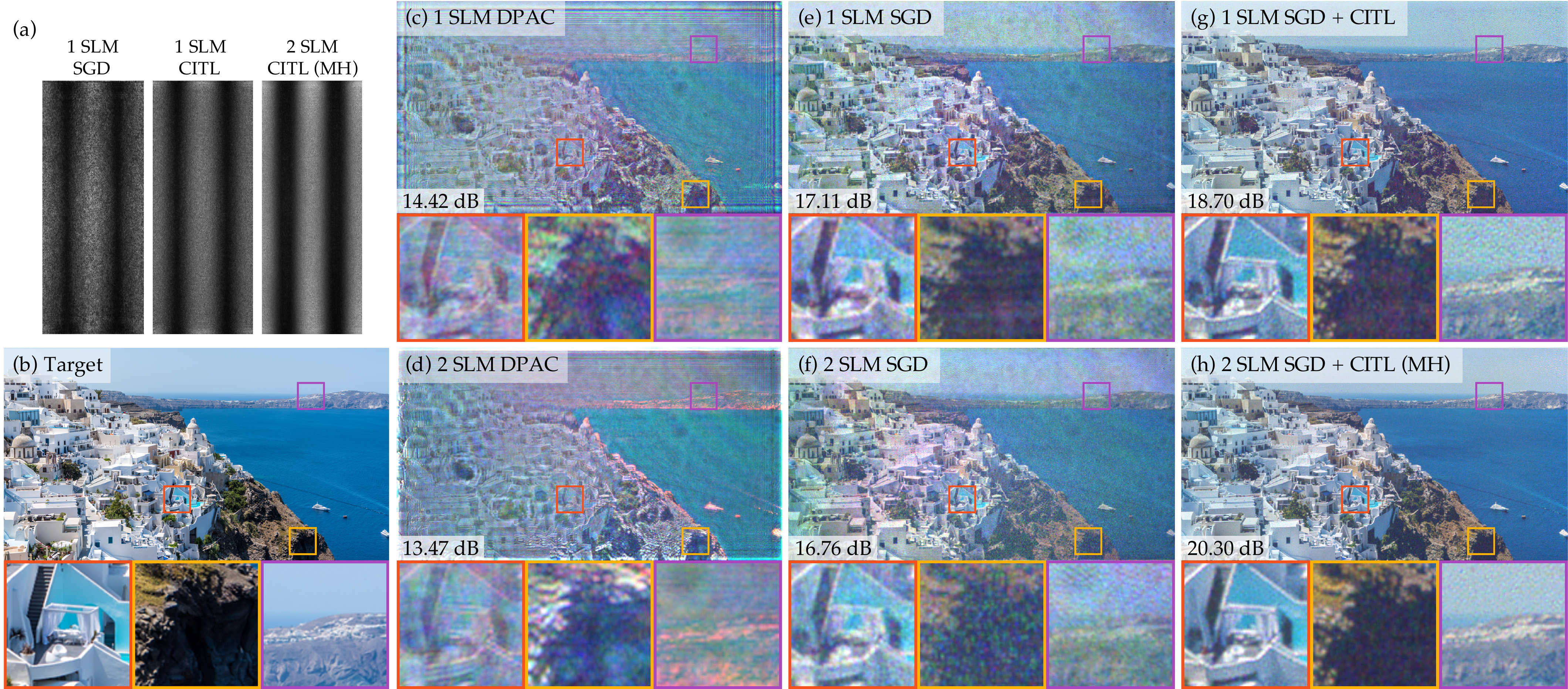}
\caption{Experimentally captured results. (a) Qualitative results of the contrast experiments shown in Table~1. SGD with \citl{} calibration achieves a higher contrast than other CGH approaches. (c)-(g) Experimental holographic images as created by several different CGH methods. Numbers indicate the PSNR w.r.t. the target image (b). Michelson Holography shows about 1.5~dB quantitative improvement over the next-best method, i.e., a single-SLM \citl{} variant of MH~\cite{Peng2020NeuralHolography},
and significant improvements in image quality, contrast, and speckle reduction compared with all other CGH methods (Visualizations~2 and~3).}
\label{fig:experimental_result}
\end{figure*}

\begin{table}[b!]
\centering
\caption{Experimental results evaluating contrast listing Weber / Michelson contrast obtained from captured sinusoidal patterns (cf. Fig.~\ref{fig:experimental_result}(a)). MH shows the the highest contrast for all wavelengths.}
\small
\begin{tabular}{cccc}
\hline
\multirow{2}{*}{Wavelength} & \multicolumn{2}{c}{1 SLM} & 2 SLM                \\ \cline{2-4} 
                            & SGD         & SGD + CITL  & SGD + CITL           \\ \hline
R (638 nm)                  & 27.3 / 0.93 & 33.4 / 0.94 & \textbf{38.2 / 0.95} \\
G (520 nm)                  & 15.5 / 0.89 & 28.4 / 0.93 & \textbf{37.0 / 0.95} \\
B (450 nm)                  & 8.75 / 0.81 & 9.51 / 0.83 & \textbf{20.9 / 0.91} \\ \hline
\end{tabular}
\label{tab:contrast}
\end{table}

Figure~\ref{fig:experimental_result}(a) and Table \ref{tab:contrast} show qualitative and quantitative experimental evaluations of contrast and image quality for several different optical and algorithmic holographic display variants. Here, Weber contrast is defined as $(I_{max}-I_{min})/{I_{min}}$ and Michelson contrast is $(I_{max}-I_{min})/(I_{max}+I_{min})$, where $I$ is the intensity of the display. Similar to our simulations, the proposed dual-SLM setup combined with \citl{} optimization achieves the best results in all cases. Especially for the blue channel, where the SLM diffraction efficiency is lower than at other wavelengths, MH shows a big improvement in contrast. The proposed method shows the uniform efficiency in the spectral domain which can correct the color shift as well.


Figures~\ref{fig:experimental_result}(c)-(h) show full-color experimental holographic images. Whereas non-\citl{} approaches, including DPAC and SGD, suffer from speckle and other artifacts, \citl{} optimization can significantly reduce these. However, using a camera in the loop allows us to automatically account for optical aberrations, phase nonlinearities, and other imperfections not directly modeled by the image formation of Eq.~(\ref{eq:asm}). Comparing the \citl-calibrated single- and dual-SLM setups demonstrates again that the proposed MH approach optimizes image quality and contrast, while reducing remaining artifacts (Visualizations~2 and~3).


\begin{figure}[h!]
\centering
\includegraphics[width=\linewidth]{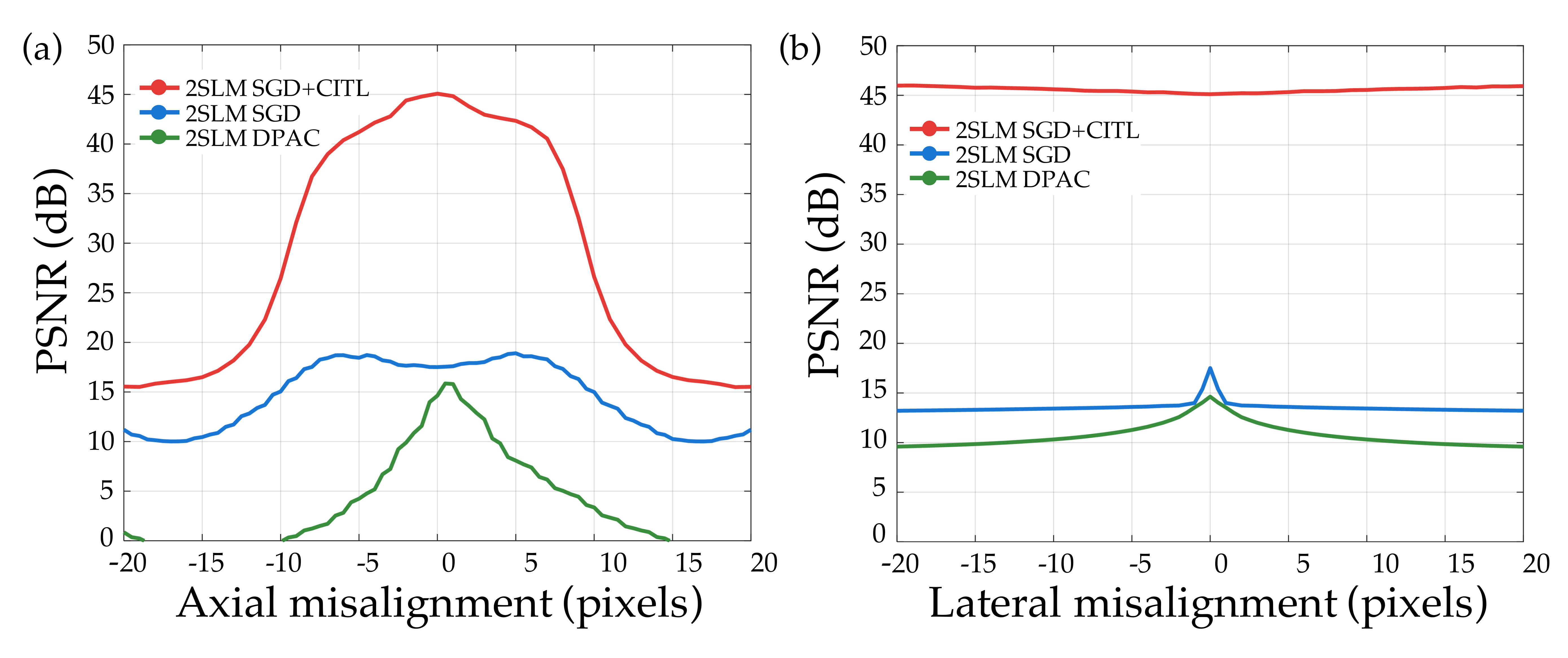}
\caption{Evaluating SLM alignment robustness in simulation. The axial and lateral misalignment between two SLMs can cause a significant image degradation, even for a misalignment on the order of the size of a single pixel. Whereas DPAC is very sensitive to small amounts of SLM alignments, SGD approaches are more robust to these with the \citl{}-calibrated SGD variant achieving by far the best and most robust results.}
\label{fig:discussion_robustness}
\end{figure}

The optical paths of the two SLMs in our setup are only roughly aligned. We primarily rely on a homography calibration of their respective contributions to the target plane. Here, the camera captures test patterns displayed on each of the SLMs and automatically derives a homography mapping between them (see Supplement for details). To study how robust MH is for slight misalignments of the SLMs, we ran several experiments in simulation for $\efficiency=0.8$. As seen in Fig.~\ref{fig:discussion_robustness}, misalignments as small as one pixel in the lateral or axial dimension result in strong image degradation for DPAC. SGD is more robust along the axial but not along the lateral dimension. SGD with \citl{} shows a similar trend for axial and lateral robustness, although with a significantly higher absolute quality. 


In summary, we demonstrate that Michelson Holography, i.e., a dual-SLM holographic display with \textsc{citl} calibration, provides superior image quality than existing CGH approaches both in simulation and experiment. 
%
Yet, our method is not without limitations. Currently, the \citl{} algorithm requires an iterative optimization process with a camera for each target image. This procedures takes around 500~seconds for a 1080p target image with our PyTorch implementation executed on an NVIDIA RTX 2080Ti. Although we could likely achieve real-time framerates using a \textsc{citl}-calibrated neural network~\cite{Peng2020NeuralHolography}, we have not attempted to do that in this letter. Our hardware prototype is currently limited by using SLMs with slightly different pixel pitches. This mismatch might cause unwanted errors and adversely affect the image quality of our results. However, it also implies that MH works robustly with different SLMs. Finally, we only demonstrate 2D holographic images in this letter and leave a detailed exploration of per-pixel 3D MH for future work.



Holographic displays are maturing into a practical technology. Artificial intelligence has enabled real-time framerates for 2D CGH~\cite{Horisaki:2018,Peng2020NeuralHolography,eybposh2020deepcgh,lee2020deep} and fully automatic display calibration for optimized image quality~\cite{Peng2020NeuralHolography,Chakravarthula:2020}. Here, we demonstrate how these neural holography techniques also enable novel holographic display hardware architectures to further improve holographic image quality, which slowly approaches that of conventional displays based on liquid crystal or organic light-emitting diode technology.


\section*{Acknowledgments}
We thank Ward Lopes, Morgan McGuire, and David Luebke for helpful discussions and advice. This project was partially supported by Ford, NSF (awards 1553333 and 1839974), a Sloan Fellowship, an Okawa Research Grant, and a PECASE by the ARO.
\newline
\newline
See Supplement 1 for supporting content.




\bibliography{ref}

\begin{thebibliography}{10}
\newcommand{\enquote}[1]{``#1''}

\bibitem{Chen:15}
J.-S. Chen and D.~Chu, {\protect\JournalTitle{Opt. Express}} \textbf{23}, 18143
  (2015).

\bibitem{Li:16}
G.~Li, D.~Lee, Y.~Jeong, J.~Cho, and B.~Lee, {\protect\JournalTitle{Opt.
  Lett.}} \textbf{41}, 2486 (2016).

\bibitem{Maimone:2017}
A.~Maimone, A.~Georgiou, and J.~S. Kollin, {\protect\JournalTitle{ACM Trans.
  Graph. (SIGGRAPH)}} \textbf{36}, 85 (2017).

\bibitem{yu2017ultrahigh}
H.~Yu, K.~Lee, J.~Park, and Y.~Park, {\protect\JournalTitle{Nature Photonics}}
  \textbf{11}, 186 (2017).

\bibitem{Shi:2017}
L.~Shi, F.-C. Huang, W.~Lopes, W.~Matusik, and D.~Luebke,
  {\protect\JournalTitle{ACM Trans. Graph. (SIGGRAPH Asia)}} \textbf{36}, 236:1
  (2017).

\bibitem{Chakravarthula:2019}
P.~Chakravarthula, Y.~Peng, J.~Kollin, H.~Fuchs, and F.~Heide,
  {\protect\JournalTitle{ACM Trans. Graph. (SIGGRAPH Asia)}} \textbf{38}
  (2019).

\bibitem{Park:2019}
J.-H. Park and M.~Askari, {\protect\JournalTitle{Optics express}} \textbf{27},
  2562 (2019).

\bibitem{Padmanaban:2019}
N.~Padmanaban, Y.~Peng, and G.~Wetzstein, {\protect\JournalTitle{ACM Trans.
  Graph. (SIGGRAPH Asia)}} \textbf{38} (2019).

\bibitem{lee2020wide}
B.~Lee, D.~Yoo, J.~Jeong, S.~Lee, D.~Lee, and B.~Lee,
  {\protect\JournalTitle{Optics Letters}} \textbf{45}, 2148 (2020).

\bibitem{kuo2020}
G.~Kuo, L.~Waller, R.~Ng, and A.~Maimone, {\protect\JournalTitle{ACM Trans.
  Graph.}} \textbf{39} (2020).

\bibitem{smalley2013anisotropic}
D.~E. Smalley, Q.~Smithwick, V.~Bove, J.~Barabas, and S.~Jolly,
  {\protect\JournalTitle{Nature}} \textbf{498}, 313 (2013).

\bibitem{ronzitti2012lcos}
E.~Ronzitti, M.~Guillon, V.~de~Sars, and V.~Emiliani,
  {\protect\JournalTitle{Optics express}} \textbf{20}, 17843 (2012).

\bibitem{cho2018dc}
J.~Cho, S.~Kim, S.~Park, B.~Lee, and H.~Kim, {\protect\JournalTitle{Opt.
  Lett.}} \textbf{43}, 3397 (2018).

\bibitem{kim2014anamorphic}
H.~Kim, C.-Y. Hwang, K.-S. Kim, J.~Roh, W.~Moon, S.~Kim, B.-R. Lee, S.~Oh, and
  J.~Hahn, {\protect\JournalTitle{Applied optics}} \textbf{53}, G139 (2014).

\bibitem{palima2007holographic}
D.~Palima and V.~R. Daria, {\protect\JournalTitle{Applied optics}} \textbf{46},
  4197 (2007).

\bibitem{liang2012suppression}
J.~Liang, S.-Y. Wu, F.~K. Fatemi, and M.~F. Becker,
  {\protect\JournalTitle{Applied optics}} \textbf{51}, 3294 (2012).

\bibitem{improso2017suppression}
W.~Improso, G.~A. Tapang, and C.~A. Saloma, in \emph{Photonics, Optics, and
  Laser Technology,}  (2017), pp. 208--214.

\bibitem{Peng2020NeuralHolography}
Y.~Peng, S.~Choi, N.~Padmanaban, and G.~Wetzstein, {\protect\JournalTitle{ACM
  Trans. Graph. (SIGGRAPH Asia)}}  (2020).

\bibitem{Chakravarthula:2020}
P.~Chakravarthula, E.~Tseng, T.~Srivastava, H.~Fuchs, and F.~Heide,
  {\protect\JournalTitle{ACM Trans. Graph. (SIGGRAPH Asia)}}  (2020).

\bibitem{Hahn:08}
J.~Hahn, H.~Kim, Y.~Lim, G.~Park, and B.~Lee, {\protect\JournalTitle{Opt.
  Express}} \textbf{16}, 12372 (2008).

\bibitem{yaracs2010multi}
F.~Yara{\c{s}}, H.~Kang, and L.~Onural, in \emph{3DTV-Conference,}  (IEEE,
  2010), pp. 1--4.

\bibitem{juday1991full}
R.~D. Juday and J.~M. Florence, in \emph{Wave Propagation and Scattering in
  Varied Media II,} , vol. 1558 (1991), pp. 499--504.

\bibitem{park2017characteristics}
S.~Park, J.~Roh, S.~Kim, J.~Park, H.~Kang, J.~Hahn, Y.~Jeon, S.~Park, and
  H.~Kim, {\protect\JournalTitle{Optics express}} \textbf{25}, 3469 (2017).

\bibitem{macfaden2017characterization}
A.~Macfaden and T.~Wilkinson, {\protect\JournalTitle{JOSA A}} \textbf{34}, 161
  (2017).

\bibitem{Hsueh:1978}
C.-K. Hsueh and A.~A. Sawchuk, {\protect\JournalTitle{Applied optics}}
  \textbf{17}, 3874 (1978).

\bibitem{Horisaki:2018}
R.~Horisaki, R.~Takagi, and J.~Tanida, {\protect\JournalTitle{Applied optics}}
  \textbf{57}, 3859 (2018).

\bibitem{eybposh2020deepcgh}
M.~H. Eybposh, N.~W. Caira, M.~Atisa, P.~Chakravarthula, and N.~C. P{\'e}gard,
  {\protect\JournalTitle{Optics Express}} \textbf{28}, 26636 (2020).

\bibitem{lee2020deep}
J.~Lee, J.~Jeong, J.~Cho, D.~Yoo, B.~Lee, and B.~Lee,
  {\protect\JournalTitle{Optics Express}} \textbf{28}, 27137 (2020).

\end{thebibliography}

\end{document}